\documentclass[11pt]{article}
\usepackage{marginnote}

\usepackage{color}
\usepackage{amsmath}
\usepackage{amsthm}
\numberwithin{equation}{section}

\usepackage{array}
\usepackage{tabularx}
\usepackage{appendix}
\usepackage{makecell}
\usepackage{graphicx}
\usepackage{amssymb}
\usepackage[T1]{fontenc}
\usepackage[utf8]{inputenc}
\usepackage[raggedright]{titlesec}
\usepackage{blindtext}
\usepackage{commath}
\usepackage{cite} 
\usepackage{caption}
\usepackage[textwidth = 7in]{geometry}
\usepackage{mathtools}
\usepackage[dvipsnames]{xcolor}
\usepackage{enumitem}
\usepackage{amsmath}
\usepackage{geometry}
\usepackage{lipsum}  
\usepackage{thmtools}
\usepackage{setspace}
\titleformat{\paragraph}[hang]{\normalfont\normalsize\bfseries}{\theparagraph}{1em}{}
\titlespacing*{\paragraph}{0pt}{3.25ex plus 1ex minus .2ex}{0.5em}
\makeatletter
\DeclareRobustCommand{\change}{%
	\@bsphack
	\leavevmode
	\color{magenta}%
	\@esphack
}
\DeclareRobustCommand{\stopchange}{%
	\@bsphack
	\normalcolor
	\@esphack
}
\makeatother

\stepcounter{secnumdepth}
\stepcounter{tocdepth}

\usepackage{hyperref}
\hypersetup{
  colorlinks   = true, 	
  urlcolor     = blue, 	
  linkcolor    = blue, 	
  citecolor   = magenta 	
}

\newcommand{\be}{\begin{equation}}
\newcommand{\ee}{\end{equation}}
\newcommand{\bea}{\begin{eqnarray}}
\newcommand{\eea}{\end{eqnarray}}
\newcommand{\bd}{\begin{displaymath}}
\newcommand{\ed}{\end{displaymath}}

\title{
	\vskip-1.5cm
Generalized Coulomb-type interaction embedded in a non-inertial cosmic string spacetime in a slow-rotation limit}

\author{
	M. Baradaran$^1$\footnote{marzieh.baradaran@uhk.cz, ORCID: \href{http://orcid.org/0000-0002-8455-9973}{0000-0002-8455-9973}}, 
	L.M. Nieto$^2$\footnote{luismiguel.nieto.calzada@uva.es, ORCID: \href{http://orcid.org/0000-0002-2849-2647}{0000-0002-2849-2647}}, 
	and 
	S. Zarrinkamar$^{2,3}$\footnote {saber.zarrinkamar@uva.es, ORCID: \href{http://orcid.org/0000-0001-9128-4624}{0000-0001-9128-4624}}
	\\  [0.7ex]
	\small
	$^1$\,Department of Physics, Faculty of Science, University of Hradec Kr\'alov\'e,\\
	\small
	Rokitansk\'eho 62, 500 03 Hradec Kr\'alov\'e, Czechia
	\\ [0.1ex]
	\small
	$^2$\,Departamento de F\'{\i}sica Te\'{o}rica, At\'{o}mica y \'{O}ptica, and Laboratory for Disruptive \\ 
		\small
		Interdisciplinary Science (LaDIS), Universidad de Valladolid, 47011 Valladolid, Spain
	\\ [0.1ex]
	\small
	$^3$\,Departament of Basic Sciences, Garmsar Branch,
	Islamic Azad University, Garmsar, Iran
}

\begin{document}
	
	\maketitle

\begin{abstract}
Motivated by the great interest in studying quantum and gravitational phenomena in a unified way, scalar bosons are considered in a cosmic string spacetime and in a non-inertial frame, with a generalized Coulomb-type interaction containing both inverse quadratic and inverse cubic corrections.
Solutions for this generalized interaction are shown for an arbitrary state in the slow rotation limit in a quasi-exact manner and a discussion is given of the structure of the problem, whose special case appears in the form of a doubly confluent Heun differential equation.
The previously known solution for the simplest case of this problem, corresponding to the ordinary Coulomb interaction, is recovered.
\end{abstract}


\section{Introduction}

The existence of cosmic strings, which are a kind of topological defect, was originally proposed by Kibble \cite{Kibble}, whose ideas have greatly influenced several fields of physics, including cosmology, quantum field theory, high-energy physics, and currently, quantum optics. 
Following the ideas of Kibble and others, several studies have been carried out on the formulation of cosmic strings \cite{Birrell, vilenkin 2000, Vilenkin PR, Matthew}.
A direct consequence of the work carried out on the cosmic string formalism is the proposal to modify the structure of the fundamental equations. 
This has led to the solution of several interesting problems.
For example, the modified equations of motion in the case of a massless scalar field for realistic cosmic string models in two- and four-dimensional cases were studied analytically using the Green's function approach by Allen and Ottewill \cite{Allen}.
On the other hand, the Klein-Gordon equation in cosmic string spacetime was investigated within a non-inertial reference frame by Motta and Bakke \cite{Motta}, while the Klein-Gordon equation in a rotating cosmic string spacetime was solved by Cunha and collaborators \cite{Cunha}  for a combination of linear and Coulomb interaction terms. 
Santos and Barros  \cite{Santos 2017, Santos 2018} obtained analytical solutions of the linear and Coulomb interactions in a non-inertial cosmic string spacetime within a generalized metric. 
In \cite{Vitoria} the cosmic string spacetime of the Klein-Gordon equation with several dislocations was considered and the associated Aharonov-Bohm and Sagnac effects were discussed.
The cosmic string formulation in the so-called $q$-deformed approach was studied in \cite{Sobhani}. 
The spin-zero version of the Duffin-Kemmer-Petiau equation with a linear interaction, i.e. the so-called Dirac oscillator, was solved in \cite{Hosseinpour}, and also this Dirac oscillator was solved exactly  for a certain form of cosmic string spacetime by Hosseinpour and collaborators \cite{Hosseinpour 2019}. Some interaction terms, including linear, Yukawa and Cornell-type terms, were solved in this framework in \cite{Deng}. 
The Fisher entropy within the Klein-Gordon framework of cosmic strings was derived by Huang et al. \cite{Huang}.
The Klein-Gordon oscillator in connection to the Kaluza–Klein theory was discussed by Ahmed \cite{Ahmed 2020}.  

On the other hand, He et al. studied the problem of a two-level system with cosmic string spacetime and extensively analyzed the corresponding entanglement problem \cite{He}. 
The quantum coherence of a weakly coupled two-level atom to a bath of fluctuating electromagnetic field within cosmic string spacetime was comprehensively analyzed in \cite{Huang 2020}. 
The effect of non-commutativity in momentum was studied for the case of the Klein-Gordon oscillator by Cuzinatto and collaborators \cite{Cuzinatto}. 
In a pair of papers, the Feshbach–Villars representation of the Klein-Gordon equation was analyzed within rotating and non-inertial cosmic string frameworks \cite{Bouzendana, Bouzendana 2023}. 
The theory of gravitational lensing in the framework of cosmic strings  was reviewed in \cite{Bulygin}. Vacuum polarization effects due to a cosmic string in an anti-de Sitter $(D+1)$-dimensional spacetime were studied analytically  in \cite{Santosa, Santosa 2024}. 
Interestingly, the concept of non-inertial frame has also  been investigated in relation to modern applications of quantum mechanics and its interface with other concepts and fields, including electrodynamics \cite {maxwel}, teleportation \cite {Alsing 2003, Alsing 2004, Chen 2019}, entanglement transfer \cite {Fuentes 2005, Alsing 2006, Fink 2017 Nat Comm, Torres 2019}, quantum encryption protocol \cite {2007 PRA}, Fisher information \cite {2014 Yao}, quantum metrology \cite {Ahmadi 2014}, and Airy beam propagation \cite {Wang 2023}.

Here, inspired by the applications of multipole expansion of Coulomb-like interactions \cite{Birrell, vilenkin 2000} such an interaction is considered within the framework of the Klein-Gordon equation when a non-inertial term is present. The structure of the present study, which also has  pedagogical value,  is as follows. In Section~\ref{sec2} we first review the Klein-Gordon equation in a cosmic string spacetime when a non-inertial terms is present. 
In Section~\ref{sec3} we will review the mathematical structure of the problem, which, as will be shown, does not appear, as far as we know, in the form of a  special function known in mathematical physics. 
Rather, it appears as a generalization of the doubly confluent Heun equation, which has not yet been addressed in the literature in closed series form. To address the equation, we propose the quasi-exact Bethe-ansatz approach. The special case of the Coulomb interaction is then revisited. 
The article concludes with some final remarks.

\section{Scalar field in cosmic string spacetime in the presence of a non-intertial term}\label{sec2}

The cosmic string is usually considered to be cylindrically symmetric and static with the metric
\be\label{primary metric}
ds^2=-dt'^2+dr'^2+\alpha^2r'^2d\phi'^2+dz'^2,
\ee
where $\alpha=1-4G\mu$ (whit $\mu$ being the the mass density of the cosmic string) denotes the form of curvature, and, obviously, $\alpha=1$ results in the case of flat spacetime. Generalization to a non-inertial frame rotating at constant velocity is possible through the transformation
\be
t'=t, \quad r'=r, \quad \phi'=\phi+\omega t, \quad z'=z, 
\ee
where $\omega$ is the angular velocity of the rotating frame. Therefore, the line element is 
\be\label {metric}
ds^2=-(1-\alpha^2r^2\omega^2)dt^2+2\alpha^2r^2\omega \,dt\; d\phi+dr^2+\alpha^2 r^2 d\phi^2+dz^2,
\ee
where $0<r<\frac{1}{\alpha\omega}$ to ensure that the particle's velocity does not exceed the speed of light. To provide an analytical equation with suitable boundary condition, i.e., solutions that vanish at the boundaries, we consider the slow-rotation limit corresponding to $\alpha \omega \ll 1  \ ( 1/\alpha \omega \rightarrow \infty)$. 
Considering the gauge  $m\rightarrow m+V(r)$, the scalar Klein-Gordon equation has the form  \cite {Birrell}
\be
-\frac{1}{\sqrt{-g}}\partial_\mu (g^{\mu \nu } \sqrt{-g}\partial_\nu \psi)+(m+V)^2\psi=0,
\ee
where $m$ is the mass of the particle and $g$ denotes the determinant of the metric tensor. We will use the method of separation of variables to find solutions of the form
\be
\label{primary ansatz}
\psi(t,r,z,\phi)=e^{-i\epsilon t}\ e^{-i\ell\phi}\ e^{ip_z z}\ R(r),
\ee
where $\ell=0,\pm 1, \pm 2, \pm 3, ...$, $p_z$ stands for the momentum components in the $z-$direction and $\epsilon$ can be regarded as the energy of the particle. Some simple algebra gives \cite {Santos 2017}
\be\label{KG}
\left\{\frac{d^2}{dr^2}+\frac{1}{r} \frac{d}{dr}-\frac{\ell^2/\alpha^2}{r^2}-(m+V(r))^2+(\epsilon+\omega \ell)^2-p_z^2\right\} R(r)=0.
\ee
Performing the transformation $R(r)=r^{-1/2}\,u(r)$, we arrive at an equation of Schr\"odinger type 
\be\label{KGnew}
\left\{\frac{d^2}{dr^2}+\frac{\alpha ^2-4 \ell ^2}{4 \alpha ^2 r^2}-(m+V(r))^2+(\epsilon+\omega \ell)^2-p_z^2\right\}u(r)=0.
\ee
Let us now introduce the following generalized Coulomb-like interaction:
\be\label{GenPot}
V(r)=\frac{a_c}{r}+\frac{a_r}{r^2}+\frac{a_q}{r^3},\qquad\text{with}\qquad a_c\,,a_r\,,a_q<0 , 
\ee
which may correspond to the far-field expansion of an axisymmetric mass distribution in cylindrical coordinates.
The terms $a_r/r^2$ and $a_q/r^3$ give post-Newtonian (relativistic) and quantum corrections, respectively.

\section{Bethe-ansatz approach for the generalized Coulomb interaction}\label{sec3}

Before addressing the essential part of this work, let us first discuss the special cases of the generalized gravitational interaction. Obviously, when only the ordinary Coulomb term is present in \eqref{GenPot}, the problem resembles the case of the non-relativistic Coulomb problem already addressed in \cite{Santos 2017} and which will finally be recovered here as a special case.
The problem when relativistic and Coulomb corrections are present resembles the so-called double-confluent Heun function, addressed by several authors using different techniques within the framework of various wave equations \cite{Ronveaux, El-Jaick, Ishkhanyan 2016, Turbiner, Turbiner88, Artemio94, BarPan18, PLB}. We then begin the analysis of the equation's structure using the Bethe-ansatz approach.

\subsection{General solution}

Substituting \eqref{GenPot} into \eqref{KGnew}, we have the Schr\"odinger-like equation
\begin{equation}\label{eqKG}
	\left\{ \frac{d^2}{dr^2}+ V_{eff}(r)\right\}u_{n}(r)=0,
\end{equation}
with an effective potential
\begin{equation}\label{veff}
	\begin{aligned}
 V_{eff}(r)=&(\omega  \ell +\epsilon_n )^2-\left(m^2+p_z^2\right)-\frac{2 a_c m}{r}-\frac{\alpha ^2 \left(4 a_c^2+8 a_r m-1\right)+4 \ell ^2}{4 \alpha ^2 r^2}  \\
& -\frac{2 (a_c a_r+a_q m)}{r^3}-\frac{2 a_c a_q+a_r^2}{r^4}-\frac{2 a_qa_r}{r^5}-\frac{a_q^2}{r^6}\,. 
 	\end{aligned}
\end{equation}
To distinguish between the various quantum states, we have added the index $n$ to $\epsilon$ and the reduced radial function $u(r)$. Then, to ensure the appropriate asymptotic behavior of the wave function at the origin and at infinity, we propose that the solution have the form
\begin{equation}\label{WavAns}
u_n(r)=\exp\left[\mathrm{A}  r+\frac{\mathrm{B}}{r}+\frac{\Gamma }{r^2}+\Delta \ln r  \right]\, \phi_n(r),
\end{equation}
where the new radial function $\phi_n(r)$ must be a polynomial  of order $n=0,1,2,...$, being $\mathrm{A}\,,\mathrm{B}\,,\Gamma<0$ and $\Delta>0$ parameters to be determined. Substituting \eqref{WavAns} into \eqref{eqKG}--\eqref{veff} yields the  following ordinary differential equation for  $\phi_n(r)$:
\begin{equation}\label{eq332}
	\left\{r^3\frac{d^2}{dr^2}+\left( 2 \mathrm{A}\, r^3+2 \Delta \, r^2-2 \mathrm{B}\, r-4 \Gamma \right)\frac{d}{dr}+\left(\Lambda_2 \,r^2+\Lambda_1\, r+\Lambda_0\right)\right\} \phi_n(r)=0,
\end{equation}
where
\begin{subequations}\label{lamdas}
	\begin{align}
		&\Lambda_2 = 2 \mathrm{A} \Delta -2 a_c\, m           \, ,\label{lamda2}\\[2pt]
		&\Lambda_1=  -2 \mathrm{A} \mathrm{B}-a_c^2-2 a_r\, m+\Delta ^2-\Delta -\frac{\ell ^2}{\alpha ^2}+\frac{1}{4}    \, ,\label{lamda1}\\[3.5pt]
		&\Lambda_0= -2 \left(2 \mathrm{A} \Gamma +a_c\,a_r+a_q\, m+\mathrm{B} (\Delta -1)\right)      ,\label{lamda0}
	\end{align}
\end{subequations}
and the parameters $\mathrm{A}\,,\mathrm{B}\,,\Gamma$ and $\Delta$ are given by
\be \label{parameters}
		\mathrm{A}=  -\sqrt{m^2+p_z^2-(\omega  \ell +\epsilon_n )^2}   , \qquad
		\mathrm{B}=  a_r  \,  ,\qquad
		\Gamma=  \frac{a_q}2  , \qquad
		\Delta=   \frac{3}{2}-a_c>0   \,.
\ee
Following the Bethe-ansatz method \cite{Zhang,Agboola}, we assume that \eqref{eq332} has polynomial solutions of degree $n$:
\begin{equation}\label{ansatZhang}
	\phi_n(r)=\! \left\{
	\begin{array}{cl}
		1, & \quad n=0, \\ 
		\displaystyle\prod_{i=1}^n (r-r_i), & \quad n=1,2,...,
	\end{array}
	\right.  
\end{equation}
where $r_i$ are unknown roots to be determined. Then,  according to Theorem~1.1 of \cite{Zhang}, the values of the coefficients $\Lambda_j$, $j=0,1,2$ in \eqref{eq332} are given by  
\begin{subequations}
	\begin{align}
		& 	\Lambda_0=  -2 \left(-n \mathrm{B}+  (\Delta +n-1)\sum_{i=1}^{n} r_i+\mathrm{A} \sum_{i=1}^{n} r_i^2\right)     ,\label{BAEgeneqs3} \\
		 &	\Lambda_1= -n (2 \Delta +n-1)-2 \mathrm{A}  \sum_{i=1}^{n} r_i      ,\qquad\qquad  
		 	\Lambda_2=-2n \mathrm{A}   ,\label{BAEgeneqs2}
	\end{align}
\end{subequations}	
in which the roots $r_1,\,r_2,\,\dots,\,r_n$ are determined by the Bethe ansatz equations
\begin{equation}\label{BAroots}
	\sum_{ j\neq i}^{n} \frac1{r_i-r_j}+\frac{\mathrm{A}\,r_i^3+\Delta\,r_i^2-\mathrm{B}\,r_i-2\Gamma}{r_i^3}=0 , \qquad i=1,2,\dots,n. 
\end{equation}
Let us now obtain the general solutions of our equation \eqref{eqKG}. Substituting $\Lambda_2$ and $\mathrm{A}$ from \eqref{lamda2} and \eqref{parameters} into the second equation of \eqref{BAEgeneqs2}, after some manipulations, the energy $\epsilon_n$ is found to be
\begin{equation}\label{nGenEnergy}
	\epsilon_n(a_c;m,\ell,\omega,p_z)= -\ell\omega \pm \sqrt{p_z^2 +m^2\left[1-\frac{a_c^2}{\left(3/2+n- a_c\right)^2}\right]}	,
\end{equation}
whenever  the expression under the square root is positive. Note that  although the energy expression \eqref{nGenEnergy} depends only on the Coulomb term $a_c$ of the potential \eqref{GenPot}, the other two potential parameters $a_r$ and $a_q$ also play an important role. In other words, \eqref{BAEgeneqs3} and the first relation in \eqref{BAEgeneqs2} result in severe restrictions on the potential parameters given explicitly by the following closed-form expressions:
\begin{subequations}\label{araq,nGen} 
	\begin{align}
		a_r&=   \frac{2(n+1) a_c+2\sqrt{m^2+p_z^2-(\omega  \ell +\epsilon_n )^2}\, (\sum_{i=1}^{n} r_i ) -(n+1)^2+{\ell ^2}/{\alpha ^2}}{2 \left(\sqrt{m^2+p_z^2-(\omega  \ell +\epsilon_n )^2}-m\right)}   ,\\[1ex]  
		a_q&=  \frac{-(2n+1)a_r-2 \sqrt{m^2+p_z^2-(\omega  \ell +\epsilon_n )^2}\; (\sum_{i=1}^{n} r_i^2) +(2 n +1-2 a_c ) (\sum_{i=1}^{n} r_i) }{2 \left(\sqrt{m^2+p_z^2-(\omega  \ell +\epsilon_n )^2}-m\right)} \,.
	\end{align}
\end{subequations}	
On the other hand, the closed form expression for the wave function associated with the energy \eqref{nGenEnergy}, from \eqref{WavAns} together with \eqref{parameters} and \eqref{ansatZhang}, is given explicitly  by 
\begin{equation}\label{WavAnsnG}
		u_n(r)=\exp\left[-\sqrt{m^2+p_z^2-(\omega  \ell +\epsilon_n )^2} \; r+\frac{a_r}{r}+\frac{a_q }{2r^2}+\left( \frac{3}{2}-a_c\right) \ln r  \right]\,  \phi_n(r).
\end{equation}
In  equations \eqref{araq,nGen} and \eqref{WavAnsnG}, the roots $r_i$ are determined by the Bethe ansatz equations
\begin{align}\label{BArootsExpli}
	&	\sum_{ j\neq i}^{n} \frac1{r_i-r_j}+\left( \frac{3}{2}-a_c\right)\frac{1}{r_i}-\frac{ a_r }{r_i^2}-\frac{a_q}{r_i^3}-\sqrt{m^2+p_z^2-(\omega  \ell +\epsilon_n )^2}=0 ,\qquad	i=1,2,...,n.
\end{align}
In summary, for a given $n$, the `general solutions' of our equation \eqref{eqKG} are given explicitly  by \eqref{nGenEnergy}-\eqref{BArootsExpli}. The exact solutions for both the ground state and the first excited state will be shown explicitly below.

\subsection{Ground state solution}\label{sect:Grn0}

For $n=0$, from \eqref{nGenEnergy} and \eqref{WavAnsnG}, the ground state energy and the corresponding wave function are given respectively  by
\begin{equation}\label{n0Energsimp}
\epsilon_0(a_c;m,\ell,\omega,p_z)= -\ell\omega \pm \sqrt{p_z^2 +m^2\left[1-\frac{a_c^2}{\left(3/2- a_c\right)^2}\right]}	 \,,
\end{equation}
and
\begin{equation}\label{WavAnsn0}
	u_0(r)=r^{ {3}/{2}-a_c }\ \exp\left[-\sqrt{m^2+p_z^2-(\omega  \ell +\epsilon_0 )^2} \; r+\frac{a_r}{r}+\frac{a_q }{2r^2} \right].
\end{equation}
Furthermore, from \eqref{araq,nGen}, the allowed values of the potential parameters, in terms of the Coulomb term $a_c$, are given by  
\be
\label{araq,n0} 
		a_r=  \frac12 \,\frac{2 a_c-1+\frac{\ell ^2}{\alpha ^2}}{ \sqrt{m^2+p_z^2-(\omega  \ell +\epsilon_0 )^2}-m}   ,\qquad\qquad 
		a_q= -\frac12\, \frac{a_r }{\sqrt{m^2+p_z^2-(\omega  \ell +\epsilon_n )^2}-m } \,.
\ee	
Note that the Bethe ansatz equation \eqref{BArootsExpli}  plays no role for the ground state case. As stated in \eqref{ansatZhang}, $\phi_0(r)\equiv1$.

\subsection{First excited state solution}
\label{sect:Grn01}
Similarly, for $n=1$, from \eqref{nGenEnergy} and \eqref{WavAnsnG}, the energy of the first excited state and its wave function are given, respectively, by
\begin{equation}\label{n1Energsimp}
\epsilon_1(a_c;m,\ell,\omega,p_z)= -\ell\omega \pm \sqrt{p_z^2 +m^2\left[1-\frac{a_c^2}{\left( 5/2- a_c\right)^2}\right]}	,
\end{equation}
and
\begin{equation}\label{WavAnsn1}
	u_1(r)=(r-r_1)\  r^{{3}/{2}-a_c }\, \exp\left[-\sqrt{m^2+p_z^2-(\omega  \ell +\epsilon_1 )^2} \; r+\frac{a_r}{r}+\frac{a_q }{2r^2}\right]\,.
\end{equation}
In this case, starting from \eqref{araq,nGen}, the potential parameters $a_r$ and $a_q$, again in terms of the Coulomb term $a_c$ as well as $r_1$, are given by the following expressions 
\begin{subequations}\label{araq,n1} 
	\begin{align}
		a_r&=  \frac12\, \frac{ 4 (a_c-1)+2\sqrt{m^2+p_z^2-(\omega  \ell +\epsilon_1 )^2}\; r_1  +\frac{\ell ^2}{\alpha ^2}}{\sqrt{m^2+p_z^2-(\omega  \ell +\epsilon_1 )^2}-m}   ,\\[1ex]  
		a_q&= -\frac12\, \frac{3a_r+2 \sqrt{m^2+p_z^2-(\omega  \ell +\epsilon_1 )^2}\;  r_1^2 +(2 a_c -3) r_1 }{ \sqrt{m^2+p_z^2-(\omega  \ell +\epsilon_1 )^2}-m } \,.
	\end{align}
\end{subequations}		
The unknown parameter $r_1$ appearing in \eqref{WavAnsn1} and \eqref{araq,n1} is determined, from \eqref{BArootsExpli}, from the following Bethe ansatz equation
\begin{equation}\label{BArootsn1}
	\sqrt{m^2+p_z^2-(\omega  \ell +\epsilon_1)^2}\;r_1^3+\left(a_c- \frac{3}{2}\right)r_1^2+ a_r\;r_1 +a_q=0 .
\end{equation}

\subsection{The special case of Coulomb interaction}
\label{sect:CoulombGrn0}

Now, let us check the special case of Coulomb interaction, i.e. when the potential parameters $a_r$ and $a_q$ in \eqref{GenPot} vanish, by comparing our previous results with Ref.~\cite{Santos 2017}. Note that due to the cumbersome structure of the potential constraints \eqref{araq,nGen} and the roots of the Bethe ansatz equations \eqref{BArootsExpli}  appearing within them, there is probably no hope of inspecting the limit for general $n$. However, we can check this explicitly for the ground state.  To this purpose, let us rewrite the ground state energy \eqref{n0Energsimp} as
\begin{equation}\label{n0eq}
	\epsilon_0= -\ell\omega \pm \sqrt{p_z^2 +m^2\left[1-\frac{a_c^2}{\left(\frac12+ \sqrt{a_c^2-2 a_c+1}\right)^2}\right]}	 \,,
\end{equation}
where we have replaced $a_c$ in the denominator of the fraction in  \eqref{n0Energsimp} by $1-\sqrt{(a_c-1)^2}$ since $a_c$ and hence $a_c-1$ are negative. On the other hand, from \eqref{araq,n0}, we see that the potential parameters $a_r$ and $a_q$ vanish whenever $a_c=\frac12-\frac{\ell ^2}{2 \alpha ^2}\;$. Then, replacing $a_c$ in \eqref{n0eq} from this condition (only the first power term) we get 
\begin{equation*}
	\epsilon_0= -\ell\omega \pm \sqrt{p_z^2 +m^2\left[1-\frac{a_c^2}{\left(\frac12+ \sqrt{a_c^2+\frac{\ell ^2}{\alpha ^2}}\right)^2}\right]}	 \,,
\end{equation*}
which is the same as Eq. (27) in Ref.~\cite{Santos 2017} for $\tilde N=0$ (note that $\eta$ appearing there is equivalent to our potential parameter $a_c$).

\section{ Conclusion}\label{sec4}
Relativistic scalar particles were investigated within a cosmic string spacetime exhibiting non-inertial effects. Based on very recent results, a generalized gravitational potential was considered, including post-Newtonian (relativistic) and quantum corrections as interaction terms.
It was briefly discussed that the resulting equation appears in a form that can be considered a generalization of the double-confluent Heun equation for which no closed-form solutions are available. 
It was noted that, to our knowledge, the equation cannot be solved using other (non-perturbative) analytical techniques, including the general form of the Lie algebra approach, the integral transforms frequently used in quantum mechanics, etc. As a result, and based on a Bethe-Ansatz approach, arbitrary state solutions were reported. After obtaining the energy spectrum, the energy shift due to the quantum and relativistic terms was expressed in \eqref{nGenEnergy} in a non-explicit manner. The special case of the ordinary gravitational potential was recovered \cite {Santos 2017}. 
The results obtained  are important because they represent¡ a simultaneous study of gravitational and quantum effects within a non-inertial cosmic string framework. We hope that this work, in conjunction with other theoretical and experimental studies in this field, will lead to a better understanding of the interface of quantum mechanics and gravity. 
Finally, it is worth mentioning that a particular difficulty of this way of approaching the problem in question is that the values of the potential parameters are quite diverse, and this will undoubtedly motivate further studies to obtain more satisfactory numerical results on the modification of the ordinary case.

\section*{Acknowledgments}
The work of MB was supported by the Czech Science Foundation, project 22-18739S.
The research of LMN and SZ is supported by Q-CAYLE (PRTRC17.11), funded by the European Union-Next Generation UE/MICIU/Plan de Recuperacion, Transformacion y Resiliencia/Junta de Castilla y Leon,   by project PID2023-148409NB-I00, funded by MICIU/AEI/10.13039/501100011033, and also by
Castilla y León  Department of Education  and the FEDER Funds (Ref. CLU-2023-1-05).
The authors also thank M.J. Lake for his comments and criticisms that helped improve this manuscript.

$~~~~$


\begin{thebibliography}{99}

\bibitem {Kibble}
T. W. K. Kibble, 
Topology of cosmic domains and strings, J. Phys. A: Math. Gen.	 \textbf{9} (1976) 1387. 

\bibitem {Birrell}
N. Birrell and P. Davies, Quantum Fields in Curved Space (Cambridge University Press, Cambridge, 1984).


\bibitem {vilenkin 2000} 
A. Vilenkin and E. P. S. Shellard, 
Cosmic Strings and Other Topological Defects (Cambridge University Press,  Cambridge, 2000).

\bibitem{Matthew}
T. Harko and M. J. Lake, 
Cosmic strings in $f(R, L_m)$ gravity, 
Eur. Phys. J. C \textbf{75} (2015) 60. 

\bibitem {Vilenkin PR}
A. Vilenkin, 
Cosmic strings and domain walls, 
Phys. Rep. \textbf{121} (1985) 263. 
 
\bibitem{Allen}
A. Allen and A.  C. Ottewill, Effects of curvature couplings for quantum fields on cosmic-string space-times, 
Phys. Rev. D \textbf{42} (1990) 2669. 

\bibitem{Motta}
H. F. Motta and K. Bakke, Noninertial effects on the ground state energy of a massive scalar field in the cosmic string spacetime, 
Phys. Rev. D \textbf{89} (2014) 027702. 



\bibitem{Cunha}
M. S. Cunha, C. R. Muniz, H. R. Christiansen, V. B. Bezerra, Relativistic Landau levels in the rotating cosmic string spacetime, 
Eur. Phys. J. C \textbf{76} (2016) 512. 

\bibitem{Santos 2017}
L. C. N. Santos and C. C. Barros Jr., Scalar bosons under the influence of noninertial effects in the cosmic string spacetime, Eur. Phys. J. C \textbf{77} (2017) 186. 

\bibitem{Santos 2018}
L. C. N. Santos and C. C. Barros Jr., Relativistic quantum motion of spin-0 particles under the influence of noninertial effects in the cosmic string spacetime, 
Eur. Phys. J. C \textbf{78} (2018) 13. 


\bibitem{Vitoria}
R. L. L. Vitória and  K. Bakke, Rotating effects on the scalar field in the cosmic string spacetime, in the spacetime with space-like dislocation and in the spacetime with a spiral dislocation, 
Eur. Phys. J. C \textbf{78} (2018) 175. 




\bibitem{Sobhani} 
H. Sobhani, H. Hassanabadi and W. S. Chung, Effects of cosmic-string framework on the thermodynamical properties of anharmonic oscillator using the ordinary statistics and the q-deformed superstatistics approaches, 
Eur. Phys. J. C \textbf{78} (2018) 106. 

\bibitem{Hosseinpour}
M. Hosseinpour, H. Hassanabadi and F. M. Andrade, The DKP oscillator with a linear interaction in the cosmic string space-time, 
Eur. Phys. J. C \textbf{78} (2018) 93. 

\bibitem{Hosseinpour 2019}
M. Hosseinpour, H. Hassanabadi and M. de Montigny, The Dirac oscillator in a spinning cosmic string spacetime, 
Eur. Phys. J. C \textbf{79} (2019) 311.


\bibitem{Deng}
L.-F. Deng, C.-Y. Longa, Z.-W. Long, and T. Xu, The generalized K-G oscillator in the cosmic string space-time, 
Eur. Phys. J. Plus \textbf{134} (2019) 355. 

\bibitem{Huang}
Z. Huang, H. Situ and Z. He, Quantum Fisher information in the cosmic string spacetime, 
Class. Quant. Grav. \textbf{37} (2020) 175002. 

\bibitem{Ahmed 2020}
F. Ahmed, The generalized Klein–Gordon oscillator in the background of cosmic string space-time with a linear potential in the Kaluza–Klein theory, 
Eur. Phys. J. C \textbf{80} (2020) 211.

\bibitem{He}
P. He, F. Yu and J. Hu, Entanglement dynamics for static two-level atoms in cosmic string spacetime, 
Eur. Phys. J. C \textbf{80} (2020) 134. 

\bibitem{Huang 2020}
Z. Huang, Quantum coherence for an atom interacting with an electromagnetic field in the background of cosmic string spacetime, 
Quantum Inf. Process.	  \textbf{19} (2020) 370.


\bibitem{Cuzinatto}
R. R. Cuzinatto, M. de Montigny and P. J. Pompeia, Non-commutativity and non-inertial effects on a scalar field in a cosmic string space-time: I. Klein–Gordon oscillator, 
Class. Quant. Grav. \textbf{39} (2022) 075006. 


\bibitem{Bouzendana}
A. Bouzenada and A. Boumali, Statistical properties of the two dimensional Feshbach–Villars oscillator (FVO) in the rotating cosmic string space–time, 
Ann. Phys. \textbf{452} (2023) 169302.


\bibitem{Bouzendana 2023}
A. Bouzenada, A. Boumali and E. O. Silva, Applications of the Klein–Gordon equation in the Feshbach–Villars representation in the non-inertial cosmic string space–time, 
Ann. Phys. \textbf{458} (2023) 169479.

\bibitem{Bulygin}
I. I. Bulygin, M. V. Sazhin and O. S. Sazhina, Theory of gravitational lensing on a curved cosmic string, 
Eur. Phys. J. C \textbf{83} (2023) 844. 

\bibitem{Santosa}
W. Oliveira dos Santos and E. R. Becerra de Mello, Vacuum polarization induced by a cosmic string and a brane in AdS spacetime, 
Eur. Phys. J. C \textbf{83} (2023) 726. 

\bibitem{Santosa 2024}
W. Oliveira dos Santos and E. R. Becerra de Mello, , Induced current by a cosmic string and a brane in high-dimensional AdS spacetime, 
Eur. Phys. J. C \textbf{84} (2024) 224. 



\bibitem {maxwel}
Y. N. Obukhov, Electrodynamics in noninertial frames, Eur. Phys. J. C \textbf{81} (2021) 919. 

\bibitem {Alsing 2003}
P. M. Alsing and G. J. Milburn,
Teleportation with a Uniformly Accelerated Partner, 
Phys. Rev. Lett. \textbf{91} (2003) 180404. 

\bibitem {Alsing 2004}
P. M. Alsing, D. McMahon and G. J. Milburn, 
Teleportation in a non-inertial frame,
J. Opt. B \textbf{6} (2004) S834. 

\bibitem {Chen 2019} 
X. Chen and K. W. C. Chan, 
Quantum teleportation of Dirac fields in noninertial frames with amplitude damping, 
Phys. Rev. A \textbf{99} (2019) 022334. 

\bibitem {Fuentes 2005}
I. Fuentes-Schuller and R. B. Mann, 
Alice Falls into a Black Hole: Entanglement in Noninertial Frames, 
Phys. Rev. Lett. \textbf{95} (2005) 120404. 


\bibitem {Alsing 2006}
P. M. Alsing, I. Fuentes-Schuller, R. B. Mann, and T. E. Tessier, Entanglement of Dirac fields in noninertial frames, 
Phys. Rev. A \textbf{74} (2006) 3232. 

\bibitem {Fink 2017 Nat Comm}
M. Fink, A. Rodriguez-Aramendia, J. Handsteiner, A. Ziarkash, F. Steinlechner, T. Scheidl, I. Fuentes, J. Pienaar, T.C. Ralph, R. Ursin,
Experimental test of photonic entanglement in accelerated reference frames, 
Nat. Comm. \textbf{8} (2017) 15304 .

\bibitem {Torres 2019}
A. J. Torres-Arenas, Q. Dong, 
G.-H. Sun, 
W.-C. Qiang, 
S.-H. Dong,
Entanglement measures of W-state in noninertial frames, 
Phys. Lett. B \textbf{789} (2019) 93. 

\bibitem {2007 PRA}
K. Brádler, 
Eavesdropping of quantum communication from a noninertial frame, 
Phys. Rev. A  \textbf{75} (2007) 022311. 



\bibitem {2014 Yao}
Y. Yao, X. Xiao, L. Ge, X. Wang, and C. Sun,
Quantum Fisher information in noninertial frames, 
Phys. Rev. A \textbf{89} (2014) 042336. 

\bibitem {Ahmadi 2014}
M. Ahmadi, D. Edward Bruschi and I. Fuentes, 
Quantum metrology for relativistic quantum fields, 
Phys. Rev. D \textbf{89} (2014) 065028. 

 \bibitem {Wang 2023}
F. Wang, D. Liu and L. Wang, 
Propagation of Airy beams in uniformly accelerated space, 
Opt. Comm. \textbf{537} (2023) 129445. 



\bibitem {Ronveaux} 
A. Ronveaux, Heun’s differential equations (Oxford University Press, New York, 1995).

\bibitem{El-Jaick}
L. El-Jaick and B. D. B. Figueiredo, Solutions for confluent and double-confluent Heun equations, 
J. Math. Phys. {\bf 49} (2008) 083508.

\bibitem {Ishkhanyan 2016}
A. M. Ishkhanyan, Schr\"odinger potentials solvable in terms of the confluent Heun function, 
Theor. Math. Phys. \textbf{188}(1) (2016) 980. 

\bibitem{Turbiner}
A.V. Turbiner, One-dimensional quasi-exactly solvable Schr\"odinger equations,  
Phys. Rep. {\bf 642} (2016) 1.

\bibitem{Turbiner88}  
A.~Turbiner, Quasi-exactly-solvable problems and sl(2) algebra,  
Commun. Math. Phys.  \textbf{118}, (1988) 467.

\bibitem{Artemio94}  
A.~Gonzalez-Lopez, N.~Kamran, and P.J.~Olver, Quasi-exact solvability,  
Contemp. Math.  \textbf{160} (1994) 113.

\bibitem {BarPan18}
M. Baradaran, and H. Panahi, Perturbed Coulomb potentials in the Klein–Gordon equation: quasi-exact solution,
Few-Body Syst. \textbf{59} (2018) 42.

\bibitem {PLB}
M. Baradaran, L. M. Nieto and S. Zarrinkamar, 
On some quantum correction to the Coulomb potential in generalized uncertainty principle approach,
Phys. Lett. B \textbf{852} (2024) 138603. 


\bibitem{Zhang}
Y.Z. Zhang, Exact polynomial solutions of second order differential equations and their applications,  
J. Phys. A: Math. Theor. \textbf{45}, (2012) 065206. 

\bibitem{Agboola}
D. Agboola, and  Y.Z. Zhang, Novel quasi-exactly solvable models with anharmonic singular potentials,  
Ann. Phys. \textbf{330} (2013) 246.









\end{thebibliography}
\end{document}